# Sphinx: a massively multiplexed fiber positioner for MSE

Scott Smedley[1a], Gabriella Baker[a], Rebecca Brown[a], James Gilbert[b], Peter Gillingham[a], Will Saunders[a], Andrew Sheinis[c], Sudharshan Venkatesan[a], Lew Waller[a]
[a]Australian Astronomical Observatory, Faculty of Science and Engineering, Macquarie University, NSW 2109, Australia; [b]ANU Research School of Astronomy and Astrophysics, Mount Stromlo Road, Weston Creek, ACT 2611, Australia; [c]Canada France Hawaii Telescope, 65-1238 Mamalahoa Hwy, Waimea, HI 96743, USA.

## ABSTRACT

In this paper we present the Australian Astronomical Observatory's concept design for Sphinx - a fiber positioner with 4,332 "spines" on a 7.77mm pitch for CFHT's Mauna Kea Spectroscopic Explorer (MSE) Telescope.

Based on the Echidna technology used with FMOS (on Subaru) and 4MOST (on VISTA), the next evolution of the tilting spine design delivers improved performance and superior allocation efficiency. Several prototypes have been constructed that demonstrate the suitability of the new design for MSE. Results of prototype testing are presented, along with an analysis of the impact of tilting spines on the overall survey efficiency.

The Sphinx fiber positioner utilizes a novel metrology system for spine position feedback. The metrology design and the careful considerations required to achieve reliable, high accuracy measurements of all fibers in a realistic telescope environment are also presented.

## 1. INTRODUCTION

This paper presents the concept design of the Sphinx[2] Fiber Positioner and Metrology System for the Mauna Kea Spectroscopic Explorer (MSE) telescope [1]. Sphinx is designed by the Australian Astronomical Observatory (AAO) and uses the tilting spine concept to achieve a practical and robust design which meets all relevant requirements. In 2017 the MSE project office down-selected Sphinx (which competed with 2 separate phi-theta designs developed by other groups) as the preferred fiber positioner design.

AAO has extensive experience developing astronomical instrumentation for other world-class observatories, including ESO, Gemini and Subaru. The AAO have developed a unique fiber-positioning technology based on a 'tilting spine' positioner in the style of FMOS-Echidna (see [2] and [3]). AAO will apply its technology, experience and successful history to one of the most complex and involved subsystems on the MSE Telescope - the Sphinx Fiber Positioner. The AAO is confident that Sphinx will be a high-performing, cost-effective and dependable part of the MSE instrument.

## 2. OVERVIEW

Sphinx comprises the following main components mounted on the MSE telescope:

- Fiber Positioner, located at prime focus
- Support electronics for the Fiber Positioner, located on the top-end ring
- Metrology system, located near the central hole of the M1 mirror

---

1 ss@aao.gov.au

2 In Greek mythology, Sphinx is the daughter of Echidna.

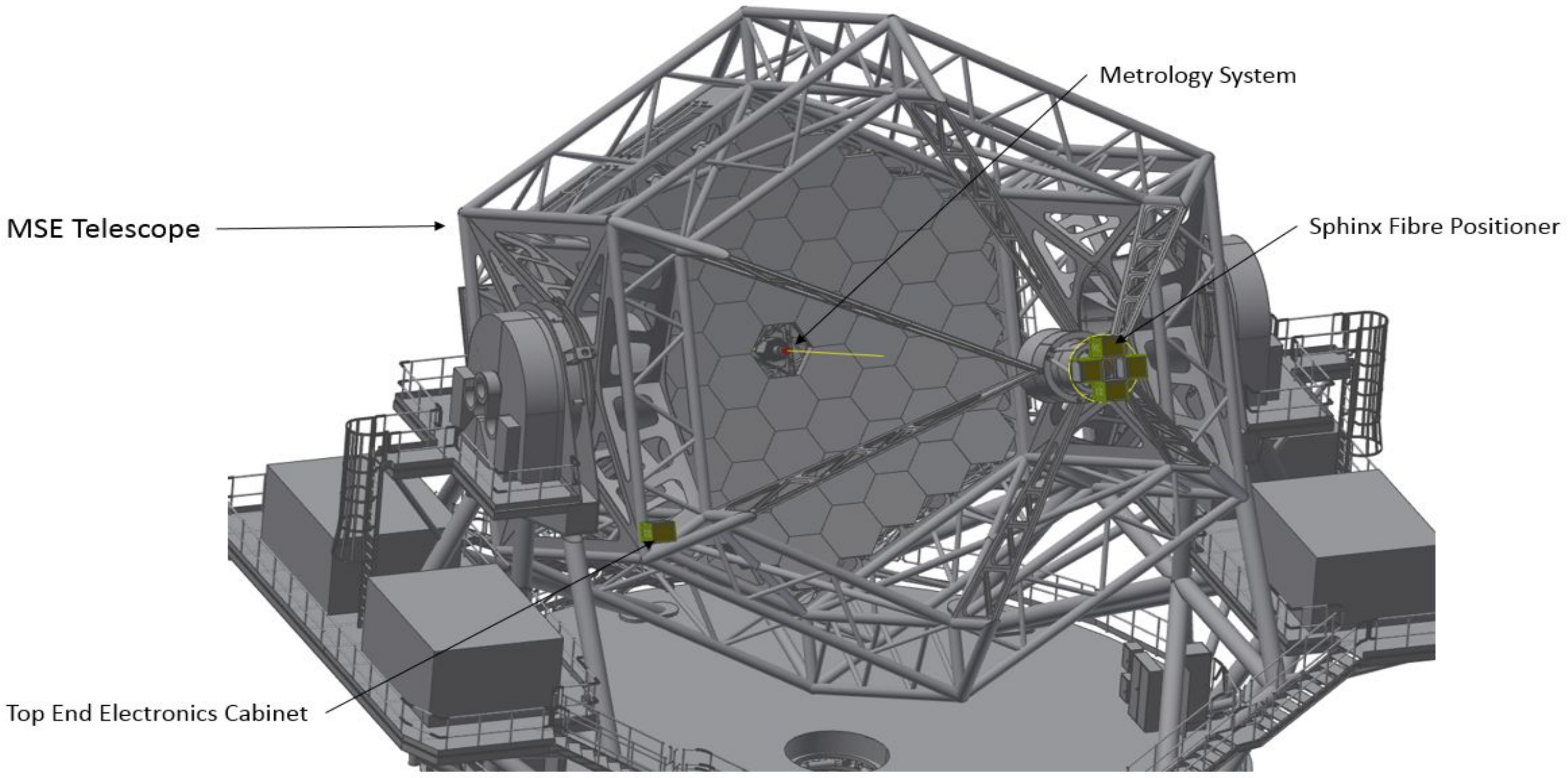

Figure 1: Sphinx comprises a Fiber Positioner (at prime focus) with support electronics (on top-end ring) and a metrology system (near the vertex of M1)

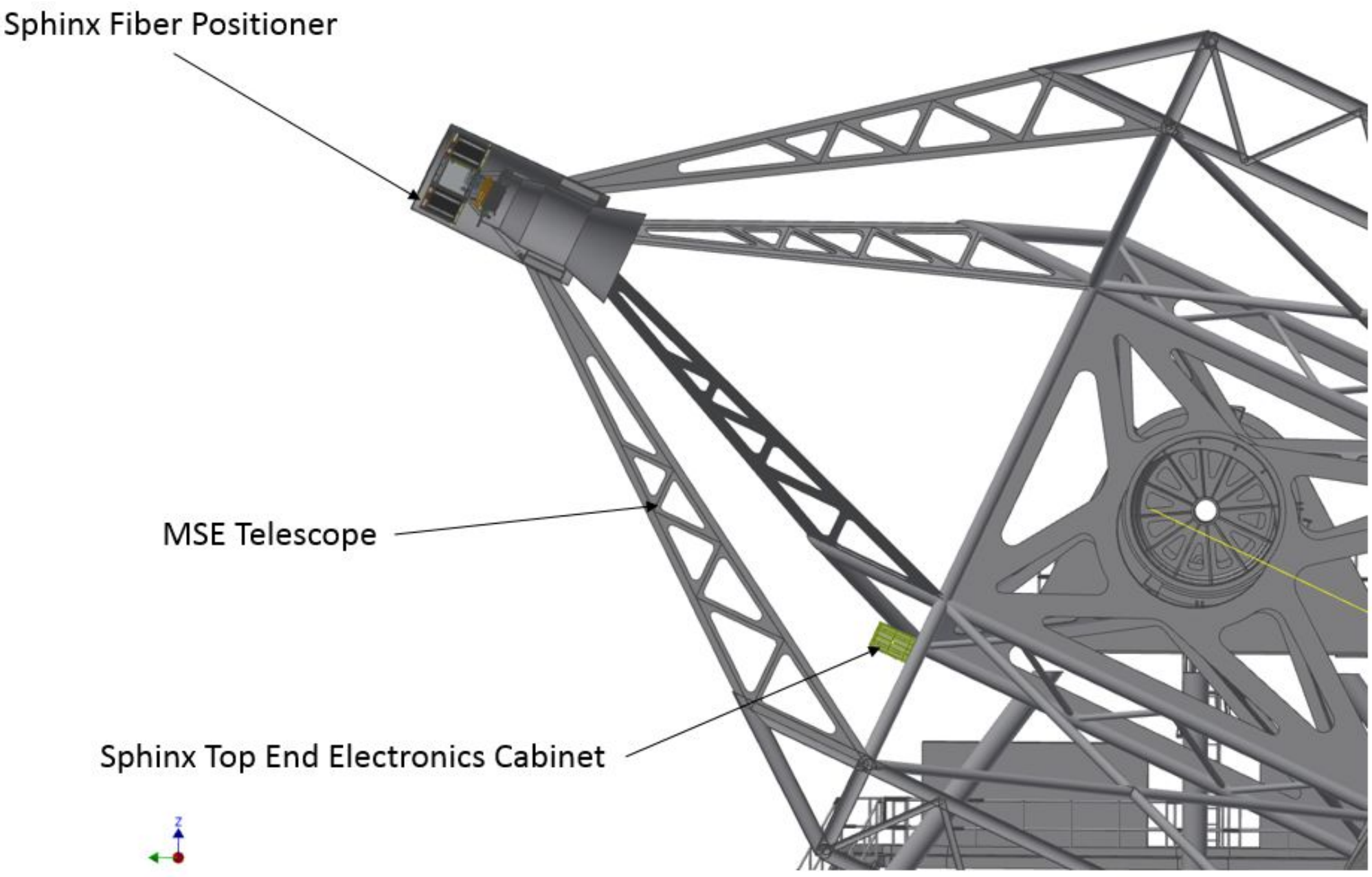

Figure 2: Fiber Positioner (at prime focus) with support electronics (on top-end ring)

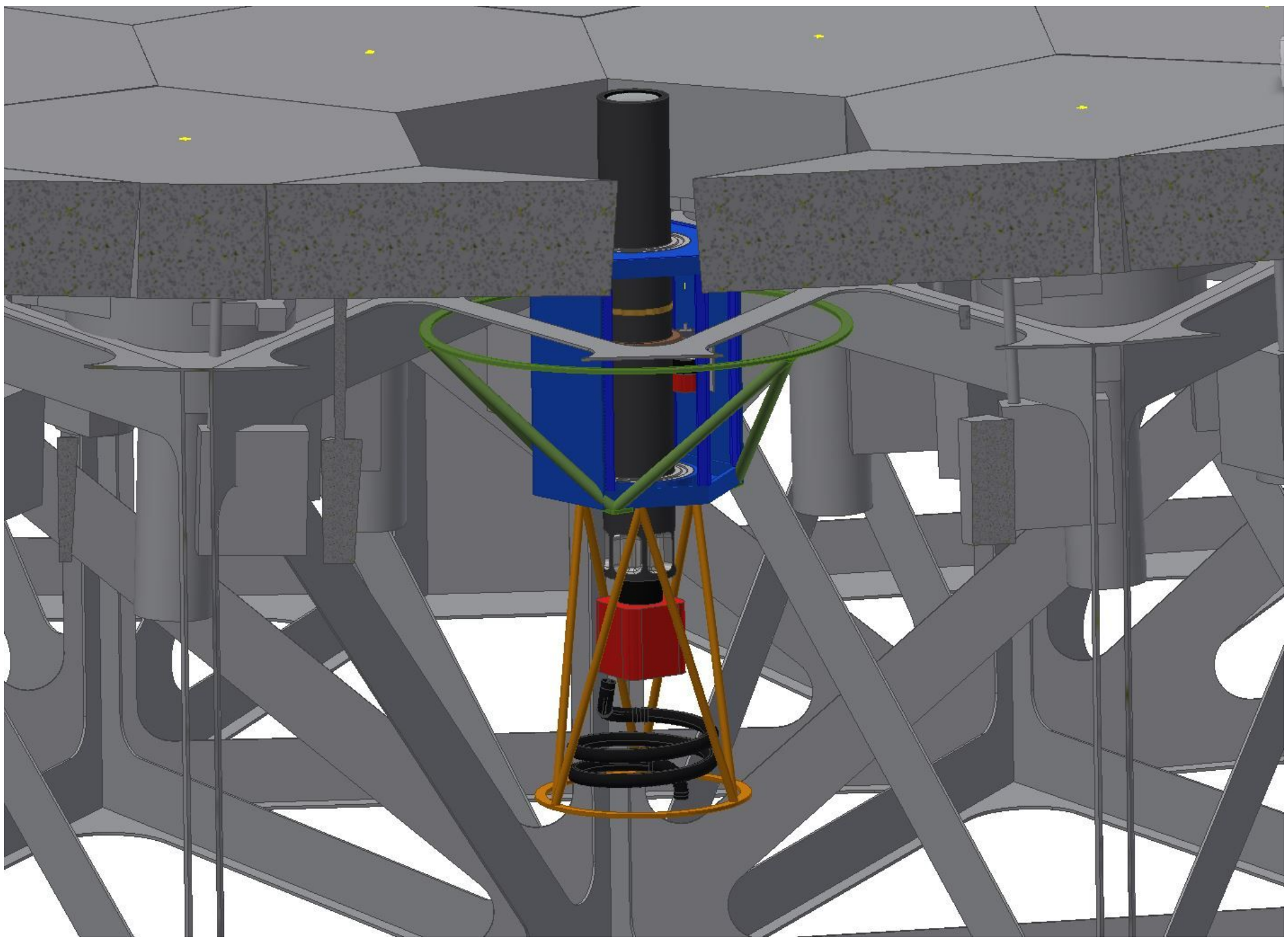

Figure 3: Location of the Metrology system (near the M1 vertex)

Sphinx consists of 4,332 piezo actuators each carrying a spine with a single science optical fiber. 1,083 fibers feed the High Resolution (HR) Spectrograph and 3,249 fibers feed the Low/Medium Resolution (LMR) spectrographs. Each actuator is capable of moving the tip of a spine within a fixed patrol radius equal to 1.24 x the pitch. The positioner deploys the optical fibers on a curved focal surface to an accuracy of 4 - 6μm RMS. Patrol areas overlap, allowing subsets of fibers to maintain full area coverage and to increase fiber allocation completeness for clustered targets, with 97% of field positions accessible by 3 or more LMR fibers and 58% of field positions accessible by 2 or more HR fibers.

Field reconfigurations are achieved in an iterative closed-loop process with positional feedback from the metrology system. Open-loop operation over small distances is also possible. Modularity and maintainability are key requirements, and as such the positioner is built from 57 modules containing 76 actuators each. The modules fit together in a triskele arrangement (Figure 10) to create a spherical surface and can be individually removed as needed.

The tilting spine concept is simple, reliable, robust and highly cost-effective. To minimise telecentricity and focus errors, the spines are manufactured as long as is practical, consistent with the available space envelope and the positioning functionality. The positioner is capable of full parallel operation of any number of spines. The hexagonal close-packed distribution of actuators across the focal surface with an actuator pitch of 7.77 mm makes this style of positioner well matched to fields with a uniform density of targets. Substantial patrol area overlap means that some targets can be reached by up to 7 different spines. Small fiber-to-fiber separations of 0.75 - 1.0 mm (7.0 - 9.4 arcseconds) and robust collision handling are also inherent features. Actuation of spines is implemented using low-cost piezoceramic elements – there are no motors or gears.

The Sphinx actuator is an evolution of the AAO's 'tilting spine' technology, first designed and implemented in FMOS-Echidna (Subaru), and later refined and simplified through design studies for WFMOS (Gemini) and WFMOS-A (AAT). The technology has also recently been developed and adopted for the AESOP positioner for 4MOST (VISTA). For Sphinx, the AAO has refined and simplified the technology, increasing performance and reliability through extensive prototyping. The Sphinx design also incorporates a novel metrology system to remotely measure the fiber locations to a high level of accuracy, coping well with dome-seeing.

## 3. FIBER POSITIONER

### 3.1 Overview

The main components of the Fiber Positioner are:

- The **main frame** carries the instrument assembly and includes the electronics assembly. 4 electronics cabinets are mounted onto the main frame in a circular pattern to fit inside a diameter of 1681mm. The frame can withstand all required loading and earthquake accelerations.
- The **base plate** interfaces with the main frame and the instrument support structure. The plate carries the mounting pads for module subassemblies and also provides for mounting of the fiber routing guides.
- The **triskele array support structure** forms the structural support for the module subassemblies. It consists of a hexagonal support structure and a Y-strut.
- The **module subassemblies** each support 76 actuators, corresponding PCB and electronic connectors. In addition, the module subassemblies carry the in-field fiducials.
- The **spines** are the rigid members carrying science and guide fibers. Each spine assembly consists of a carbon fiber tubular structure, a counterweight and a pivot ball and a ferrule through which the optical fiber is mounted.
- The **fiber routing** system ensures that the optical fibers are carried in an organised fashion from the spines to interface with the main optical fiber system which leads to the spectrographs. The fiber routing system ensures that the requirements for the minimum bend radius for the fibers are maintained.
- The **electronics cabinets** are mounted in a circular pattern which will be connected to the instrument rotator. An additional electronics cabinet will also be installed on the top end ring of the MSE telescope.

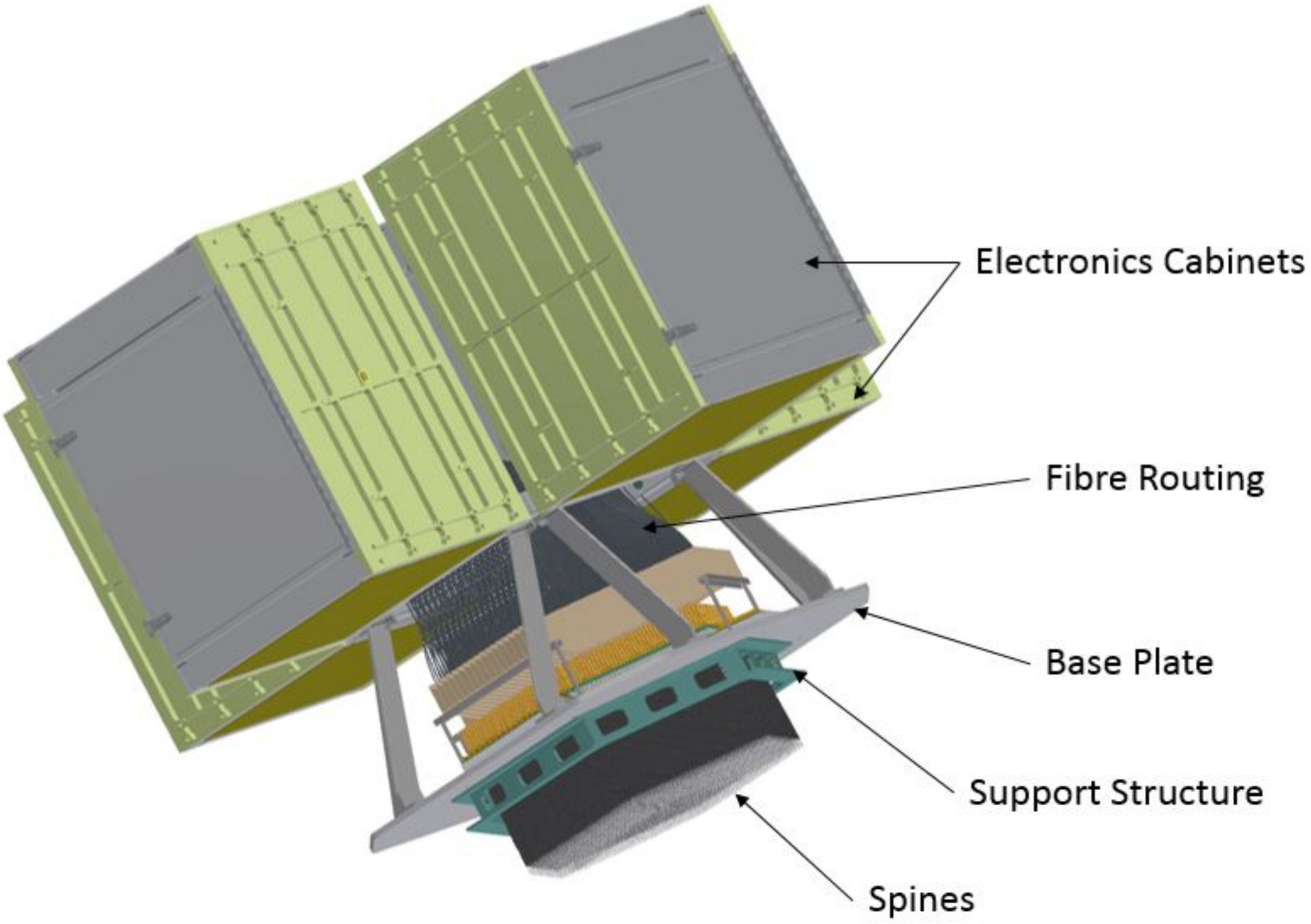

Figure 4: Main components of the Sphinx Fiber Positioner

### 3.2 Actuator

The tilting spine actuator design is an improved version of the AESOP actuator being built by AAO for ESO's 4MOST instrument (for its VISTA telescope). The actuator is comprised of 2 main components: the mount and the spine. Details about the operation of tilting spine actuators can be found in [4] and [5].

#### 3.2.1 Mount

The actuator mount consists of a cylindrical piezo ceramic tube with outer electrodes and a magnet cup assembly in which a science spine is mounted (see Figure 5).

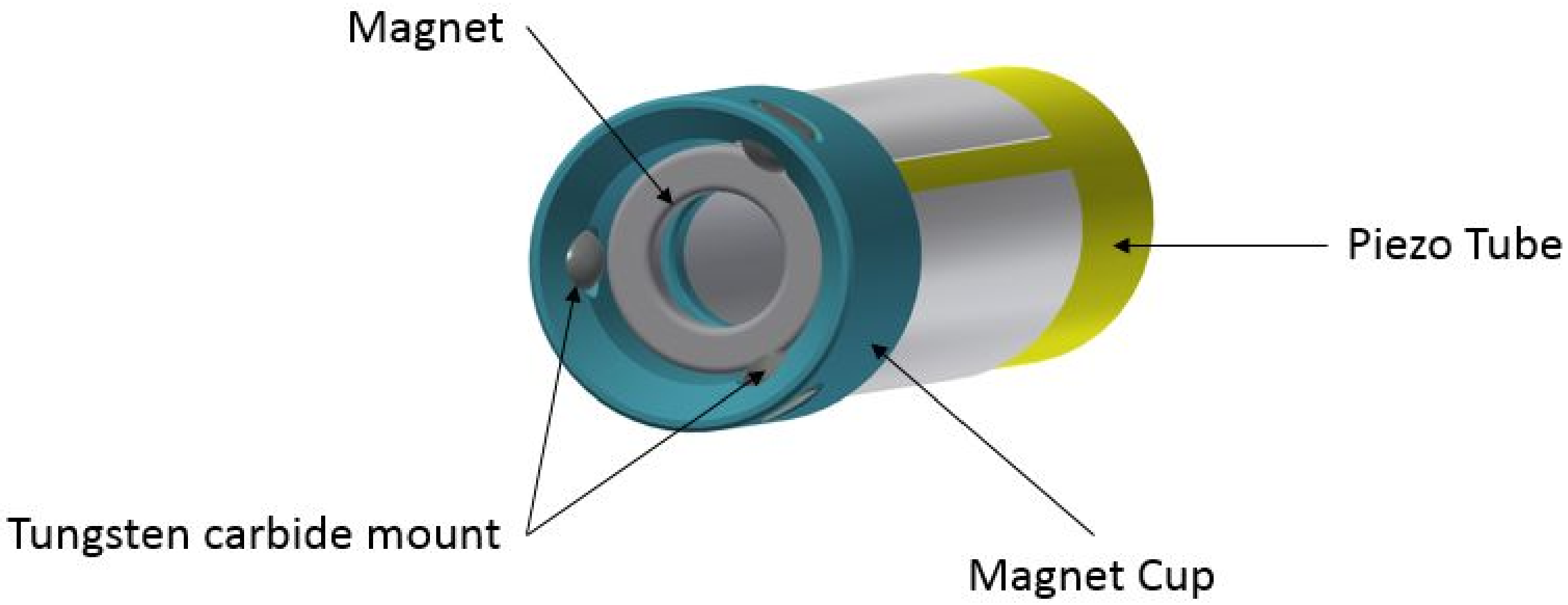

Figure 5: Actuator mount

The design of the actuator mount allows for the spine to be positioned in any direction and its step size is set by the amplitude of drive signal. Large step sizes are achieved using relatively higher voltages and smaller step sizes at lower voltages. The frequency of the drive waveform is set to avoid resonance of the spine and structure. The frequency is also optimised to control the movement speed. All spines can be moved simultaneously by providing waveforms to all actuators in parallel.

#### 3.2.2 Spine

The spine assembly consists of the main carbon fiber tube, a smaller carbon fiber tube which is installed using a telescopic arrangement, a ferrule in which optical fiber is installed, a pivot ball and a counterweight. (see Figure 6.) The spine is designed to be balanced around the pivot point and is installed into a magnet cup subassembly. The spine remains in contact with the magnet cup assembly through magnetic attraction to its kinematic mount.

The spines are located on the focal surface through modules which are mounted on a Y-strut to form an array in a triskele pattern. The length of the spine is 250mm from the pivot point to the tip of the spine. The length provides for a repeatable step size at the spine tip and reduces the optical losses due to change in focus and misalignment with the axis of the beam from the telescope. The spines use a two stage carbon fiber telescopic tubing to achieve high stiffness. The fiber is attached to the spine using a zirconia ferrule. The ferrule provides a rigid structure for the fiber tip. The fiber runs along the length of the spine and exits through the counterweight. The counterweight ensures the spine is balanced and the center of gravity is located at the pivot point.

The spine design has been carefully optimized to allow adjacent spine tips to come into very close proximity, favoring simultaneous observation of clustered targets. Spines physically touch at the spine tips at a separation of ~0.7 mm (determined by the ferrule diameter). A small additional buffer of ~50 µm is enforced by the control software, to allow some room to maneuver spines and to account for any decentering of fibers within ferrules, making the minimum separation between any 2 targets 0.75 mm (7.0 arcseconds). All spines are of the same design, so there is **no** difference in the minimum separation between HR-HR, LMR-LMR or LMR-HR fibers.

The minimum circular diameter between any 3 adjacent fibers is 1.06 mm (9.9 arcseconds).

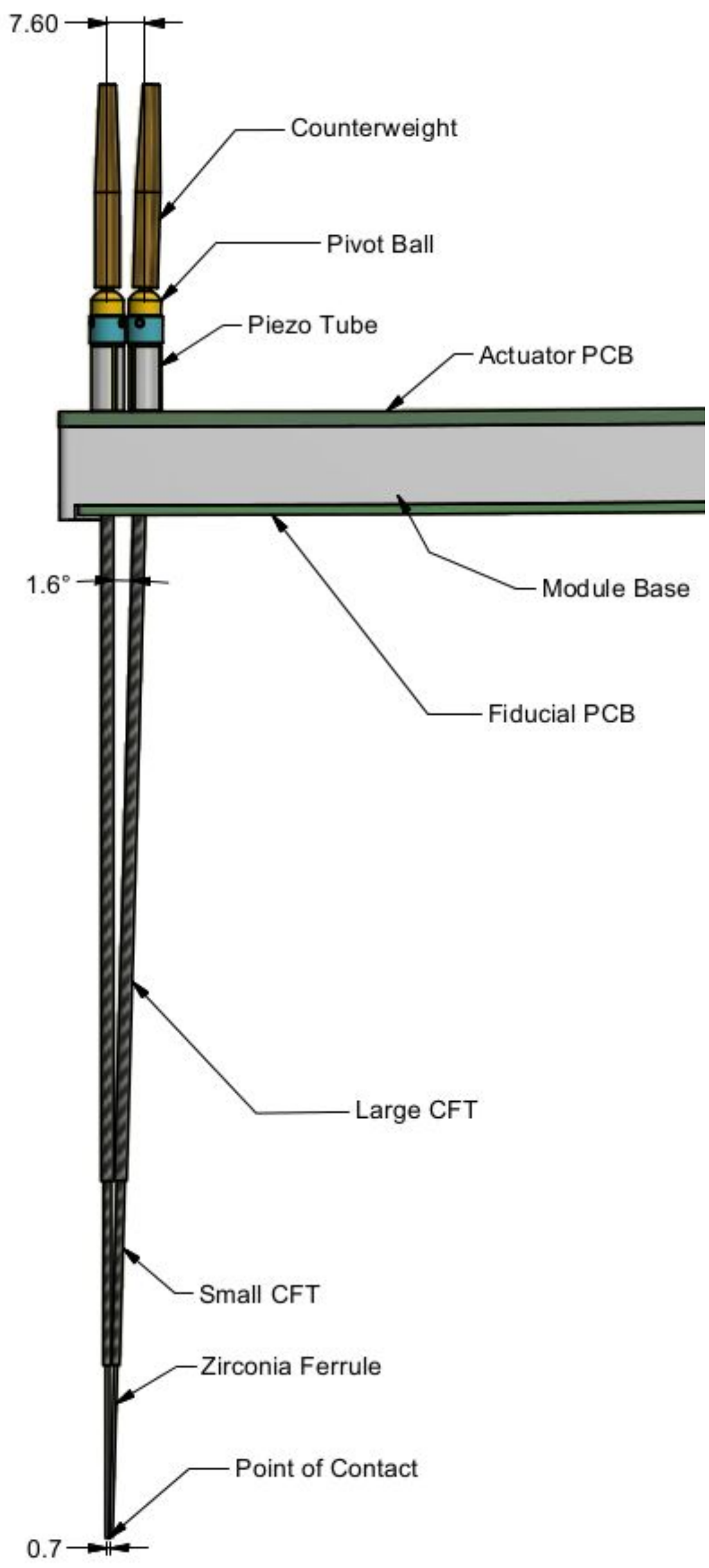

Figure 6: Proximity of 2 adjacent spine tips

### 3.2.3 Advantages

The 'Echidna' tilting spine technology is particularly suitable for high fiber densities in very fast beams, features shared by MSE. The technology is proven and mature, with 8 years successful operation for the original FMOS-Echidna, and now in Final Design phase for the AESOP positioner for 4MOST.

The radical differences between the tilting spine and phi-theta designs have significant science impacts. The relative technical advantages of the tilting spine design for MSE are summarised below.

Table 1: Advantages of tilting spine actuators

| Feature | Description |
|---|---|
| Large patrol radius | Spines have a large patrol radius, leading to multiple overlaps between the patrol areas of adjacent spines, and allowing full field coverage for subsets of the spines. This increases allocation efficiency, and/or allows the possibility of separate fiber types in separate spines. |
| Small exclusion radius | Spines have a very small exclusion radius around each target (limited by the ferrule diameter, 0.7mm), allowing very close pairs of targets to be observed. When combined with (a), this has a dramatic effect on the allocation efficiency for clustered targets. |
| Simultaneous LMR/HR full-area coverage with all fibers | Spines can have a small pitch, which when combined with (a), means the HR and LMR fibers can be carried on separate spines, with a full complement of each type of fiber, and with full field coverage for each mode. This avoids losses associated with optical switches, and also allows simultaneous HR/LMR observing with all fibers at all times, with full areal coverage for each mode. |
| Actuator fiber handling | The actuator action is extremely gentle on the fiber, with no twisting and an integrated bending of 6.5° or less, with a radius of curvature >500mm; adding negligible Focal Ratio Degradation (FRD) within the fiber itself. |
| Reliability and robustness | The AAO has an unparalleled record for designing and delivering fiber positioners, and the tilting spine technology is intrinsically very reliable, with very few moving (or even non-moving) parts, almost no maintenance requirements, insensitivity to thermal expansion effects, and immunity to damage from collisions. For all these reasons, tilting spines offer a very low technical risk. The Echidna system has shown robust operation in the field on Mauna Kea. In 8 years of Echidna operation at Subaru, only one spine failed, this corresponds to a failure rate of 0.031% per year. This high reliability means Sphinx will deliver very high availability. |
| Collision-resistant | Collisions between spines should not occur in normal operation. However in the case of a collision occurring under any (abnormal) situation, no damage occurs to the spine or the fiber. The only adverse effect is the possibility of spine identification confusion, which is easily remedied by calibration. |
| Compatibility with in-field fiducials | The tilting spine design allows large numbers of in-field fiducials – essential for accurate positioning in the presence of dome seeing - without sacrificing any science fiber feeds. |

#### 3.2.4 Prototyping

During the concept design phase, several different actuator designs were developed. Many prototypes were manufactured and tested to assess the performance of the design and refine it further. Below we present the results from the AESOP prototype, which comprised 64 spines in an 8x8 array. The actuator design utilised in the AESOP prototype is similar, but not identical, to the design proposed for Sphinx. (eg. the AESOP prototype used an actuator pitch of 10 mm, whereas Sphinx will use 7.7 mm.) The performance of the Sphinx spine actuators is expected to be no worse than the AESOP actuators, and there is good reason to believe it will be significantly better in many aspects.

##### 3.2.4.1 Positioning accuracy

The AESOP prototype was utilised to simultaneously position all 64 spines on pseudo-random targets within an 11.5 mm (1.2p) patrol radius. Spine positions were adjusted iteratively until they were within the required position tolerance (5 µm or 10 µm RMS). The total number of iterations (with the spines starting near their home positions) was recorded, and tests performed many times, each with a different set of pseudo-random targets.

Table 2: Closed loop positioning accuracy

| Accuracy requirement (um) | Final RMS accuracy (um) | Number of compound moves (avg. ± σ) | Number of Targets |
|---|---|---|---|
| 10 | 6.7 | 4.0 ± 0.2 | 146,002 |
| 5 | 4.3 | 4.8 ± 0.5 | 163,933 |

These results show the fiber positioner has no difficulty in positioning fibers on target to a tolerance of < 5 µm RMS.

#### 3.2.4.2 Distribution of positioning errors

The typical distribution of final positioning errors is lognormal-ish, as shown in Figure 7. The separated X & Y errors closely follow a normal distribution.

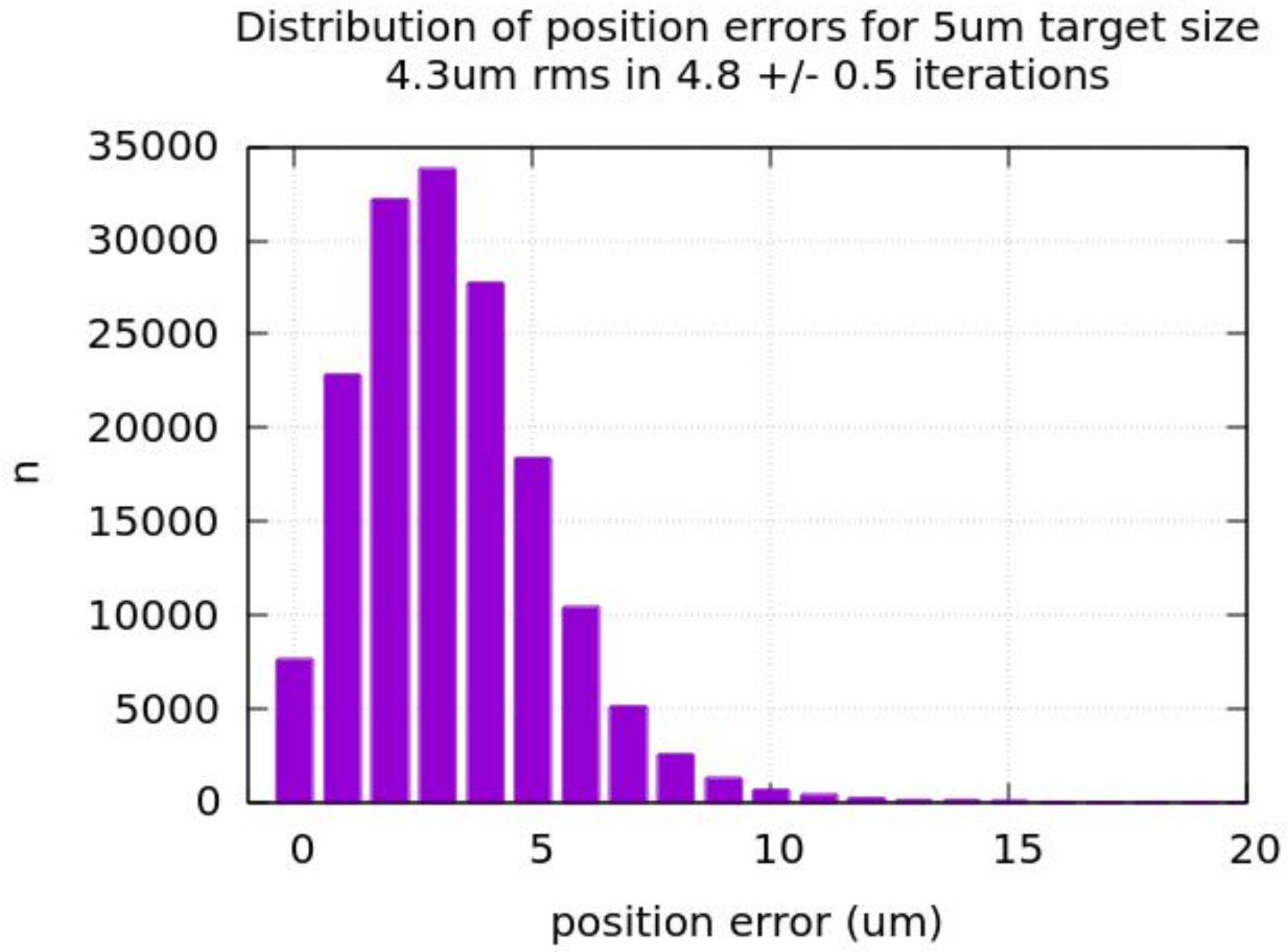

Figure 7: Distribution of final positioning errors

#### 3.2.4.3 Durability

The AESOP prototype was operated almost continuously for several months to test the durability of the spine design, repeatedly positioning each active spine on a pseudo-random target. After traversing >3.0 km (>64% the maximum distance spines are expected to travel in their lifetime for MSE), none of the spines exhibited any significant deterioration in performance.

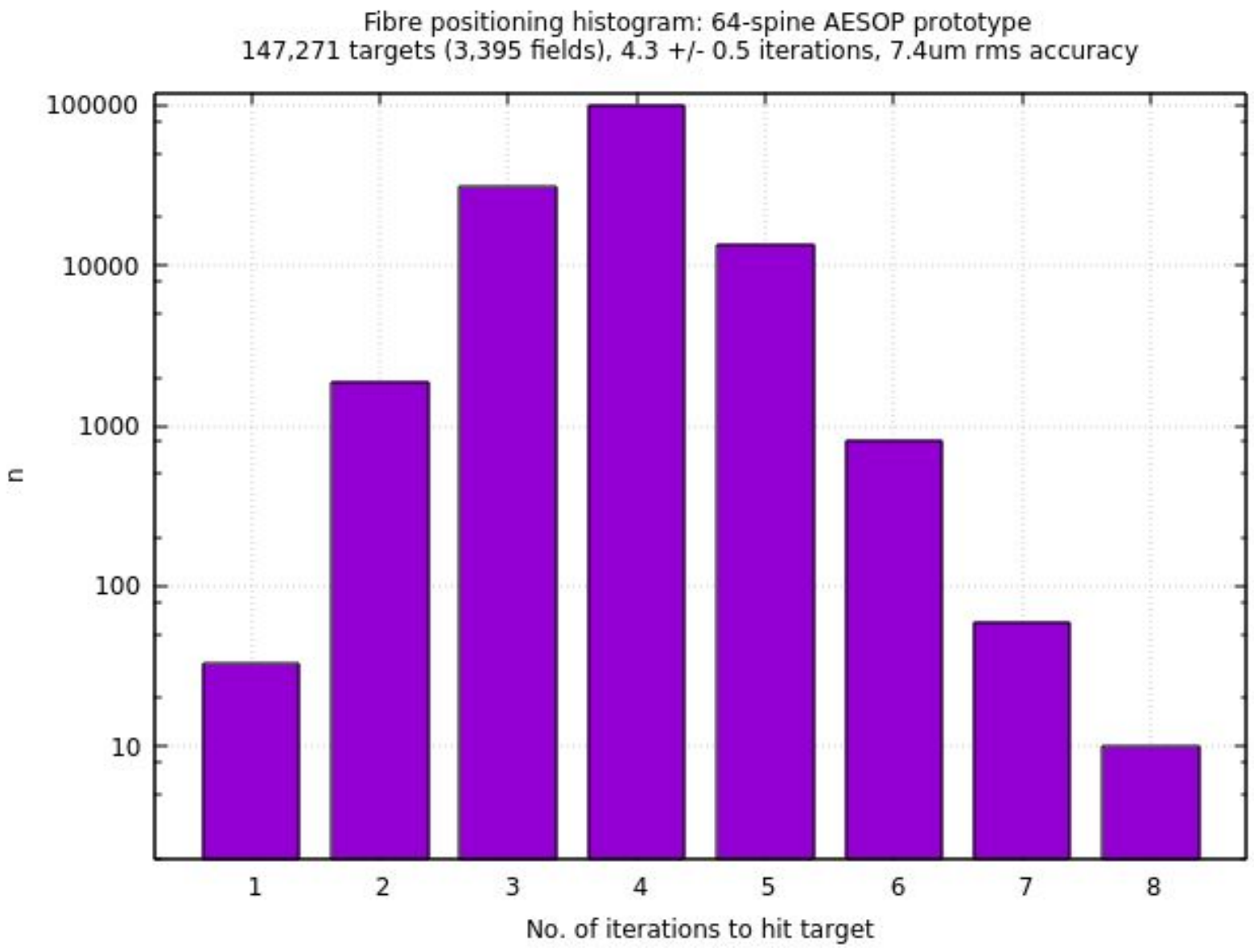

Figure 8: Performance of the AESOP prototype spines after traversing >3.0 km

AESOP spines continued their excellent performance long after they had exceeded the AESOP expected lifetime. This reinforces our confidence that the spine and actuator design is robust and unlikely to deteriorate during 20 years of operation for MSE.

#### 3.2.4.4 Environmental testing

MSE requires that its Fiber Positioner function between -12°C and +26°C, and fully meet all of its performance requirements between -3°C and +12°C. To verify compliance with these requirements, 2 AESOP prototype modules were installed in a climate chamber at AAO and pseudo-random positioning tests were conducted at various temperatures and the positioning performance quantified.

Table 3: Spine performance vs temperature

| Temperature (°C) | RMS moves per spine | Number of iterations (avg. ± σ) | Number of Targets |
|---|---|---|---|
| -10 | 4.3 | 5.1 ± 0.6 | 174 |
| 0 | 3.9 | 4.5 ± 0.9 | 1,715 |
| 5 | 4.2 | 5.1 ± 1.3 | 3,903 |
| 10 | 4.6 | 5.8 ± 1.3 | 1,120 |
| 20 | 4.9 | 5.9 ± 1.0 | 3,879 |

The spine calibration model was not configured to account for temperature variations in step size, though this functionality is intended to be used with the final instrument. Even with this artificial restriction, the results of temperature testing show that there is no difficulty in positioning spines on target, within tolerance, at temperatures to as low as -10°C.

### 3.2.5 A new low voltage actuator

The preferred actuator design for Sphinx utilises a slightly modified piezo tube to permit low voltage operation. At ±15V this low voltage Sphinx actuator yields, in theory, an identical deformation to the AESOP actuator operating at ±150V. In addition to improved safety, a key advantage of the low voltage actuation method is the ability to control a single piezo electrode independent of all other electrodes, to achieve omni-directional control. Both closed-loop and open-loop accuracy of the low voltage actuator is expected to be better than high voltage AESOP actuator, largely due to the dedicated drive electronics for each actuator which permit omni-directional movement (spines can move in any direction directly toward their target, reducing the travel distance and move error by a factor of up to $\sqrt{2}/2$) and the elimination of the discreteness error (where the drive voltage can be dynamically adjusted to give an optimal step size). The new low voltage piezo tube actuator design will be prototyped during the next phase of the project.

### 3.3 Module assembly

To aid assembly and maintenance, spines are mounted on identical modules that are removable from the main positioner assembly. The positioner assembly consists of a triskele array of 57 modules. Each module carries 76 spines. The modules are mounted on the Y-strut, which forms the support structure. An electrical connector is mounted on the outer end of the module. A full module assembly is shown in Figure 9.

The module base is 415 mm long and forms a structural base on which the actuators are mounted. Two printed circuit boards are mounted on the module base, one circuit board powering each of the actuators and the other mounted on the bottom of the module base carrying the fiducial LEDs and the corresponding light shield.

Holes on the module base accommodate the piezo tubes and allow the spines to pass through whilst allowing sufficient patrol radius for each spine. Fiber routing is provided for each of the modules but is supported by the main assembly frame. This allows for individual spines to be removed without the actuators themselves being removed. For further maintenance in the unlikely event of actuators requiring replacement, each module can be replaced as a subassembly.

To place fibers on the spherical focal surface, identical modules using the same curvature are used. Each module is assembled with a precise jig to ensure all the pivot balls are assembled to a spherical surface with a precise offset from the field of view.

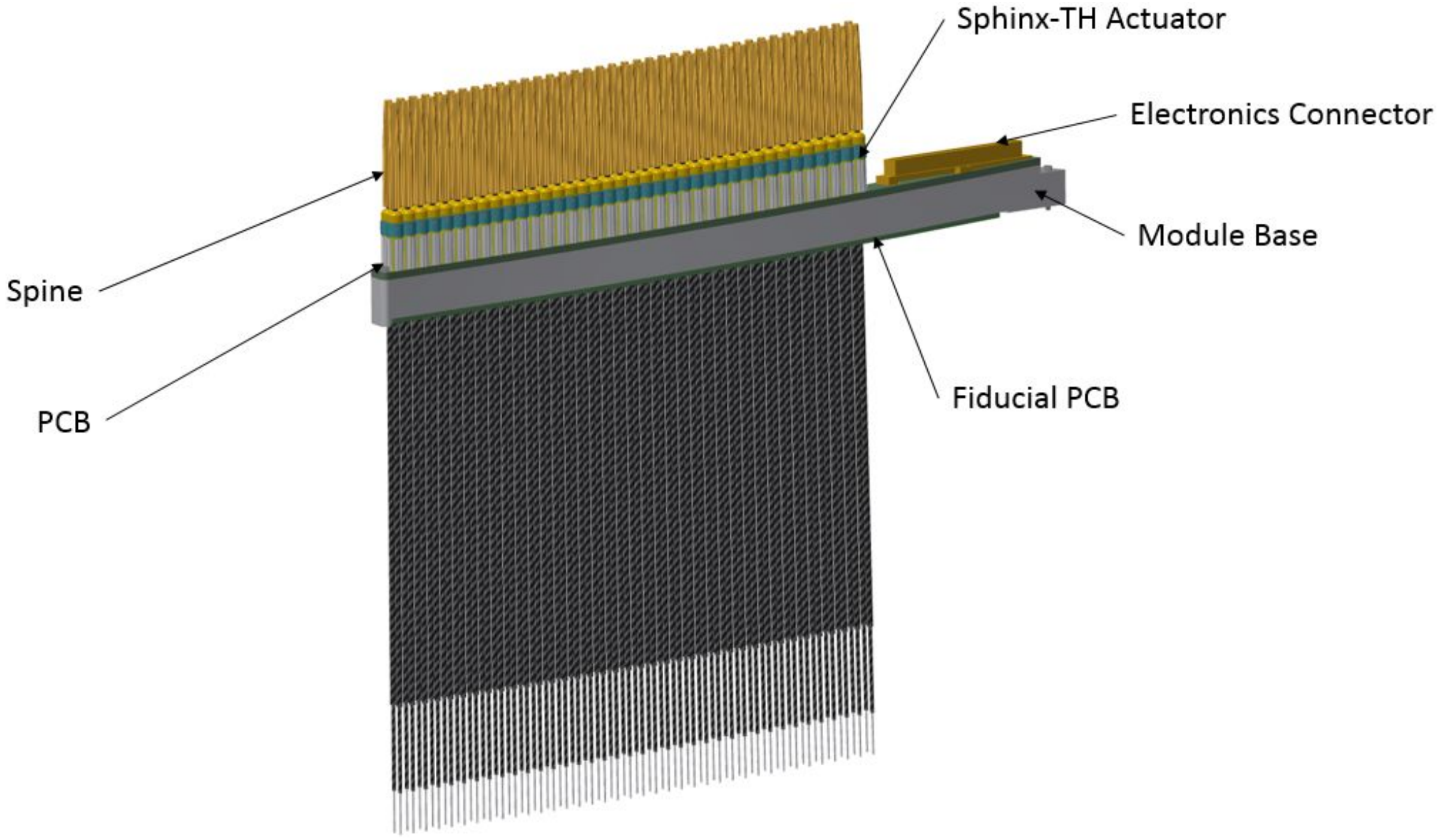

Figure 9: Full module assembly carrying 76 actuators, electrical connector and PCB

### 3.4 Support structure assembly

The Fiber Positioner support structure consists of the base plate, module mount pads, Y-strut and the hexagonal support structure. The Y-strut and the base plate form the primary support and mounting for the triskele module arrangement.

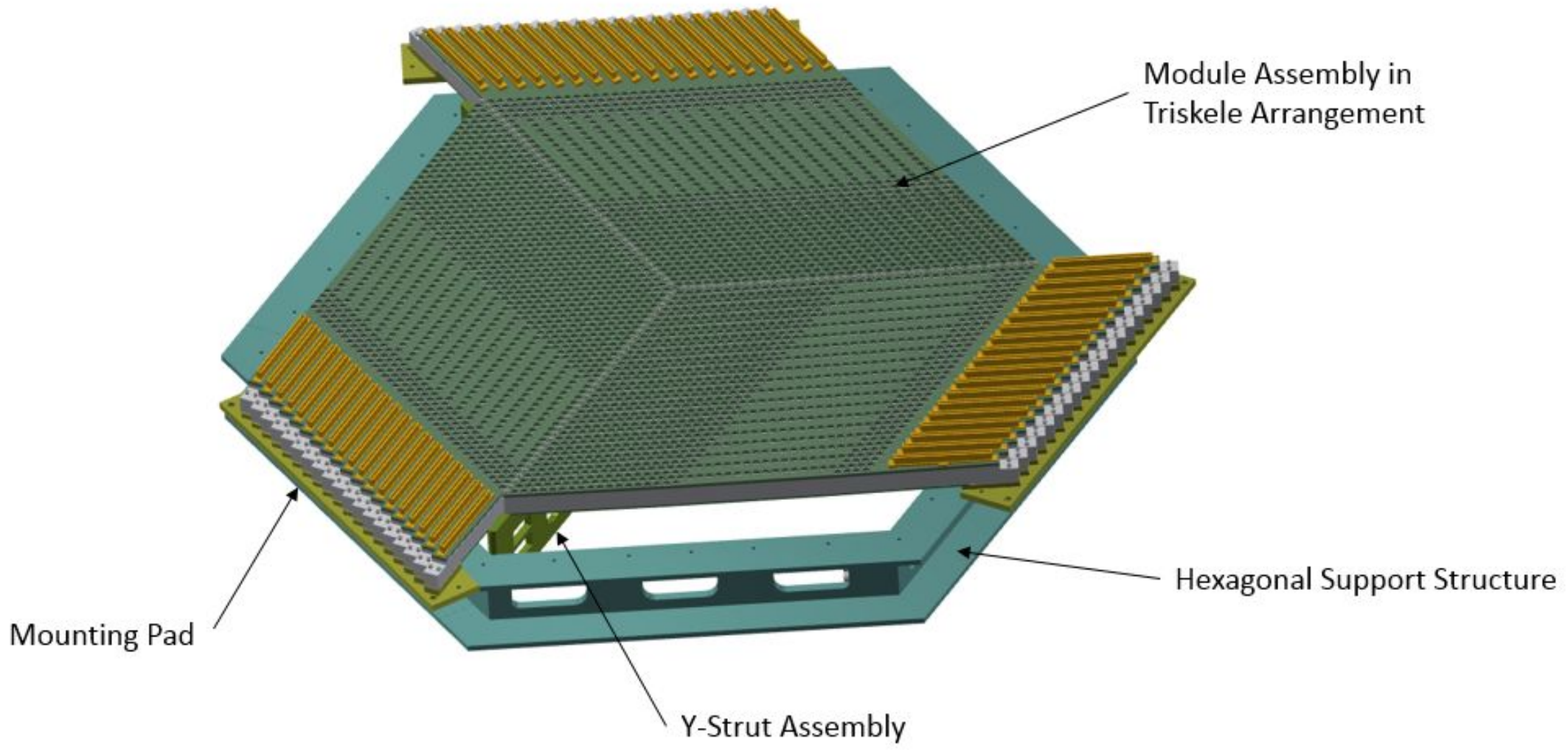

Figure 10: Modules in Triskele arrangement, mounting pads and support structure

### 3.5 Focal surface

#### 3.5.1 Layout

The Sphinx focal surface is a hexagonal section of a spherical surface (of radius 11,325 mm). Figure 11 shows the spines in the triskele arrangement at their respective home positions. The spine tips provide full field coverage of the entire hexagonal focal surface, which measures 584 mm corner-to-corner. The focal surface is convex as viewed from the primary mirror vertex.

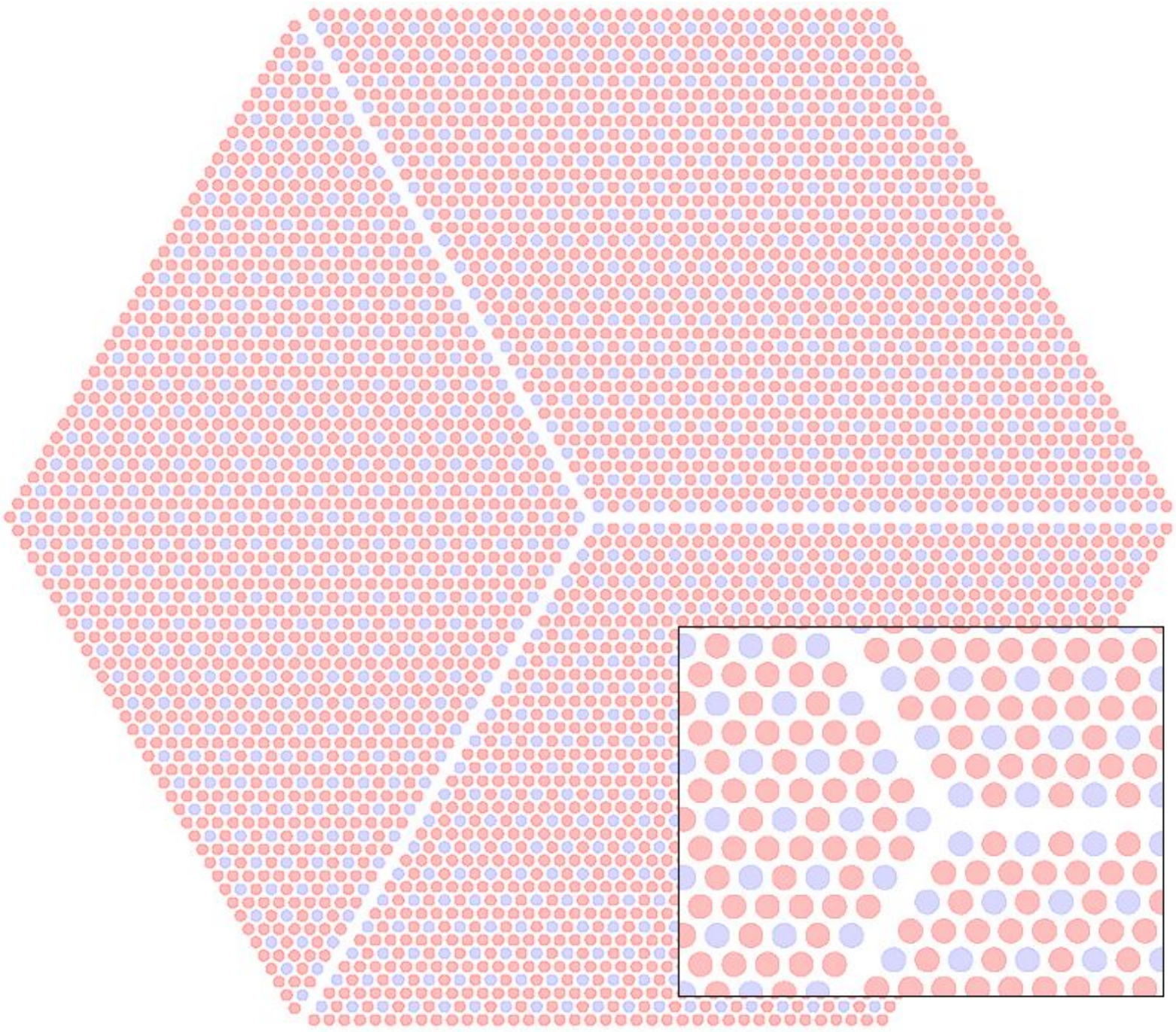

Figure 11: Focal surface actuator arrangement: LMR (red) and HR (blue)

### 3.5.2 Field coverage

Due to the large patrol radius of its spines (1.24*p*, where *p* is the separation between spine tips at their home positions), the Sphinx Fiber Positioner offers excellent field coverage for both the LMR and HR fibers. The total field coverage (aka fill factor) for LMR & HR fibers is shown in Table 4. The Y-strut support structure creates a discontinuity in the otherwise uniform coverage (see Figure 11, Figure 12, and Figure 13). This is due to an 11.35 mm separation of spine home positions across the strut, compared to the nominal 7.77 mm spine tip pitch.

Table 4: LMR and HR field coverage

| Number of fibers | Field coverage | |
|---|---|---|
| | LMR | HR |
| 1 or more | 99.99% | 100.00% |
| 2 or more | 99.83% | 58.06% |
| 3 or more | 97.07% | 4.47% |

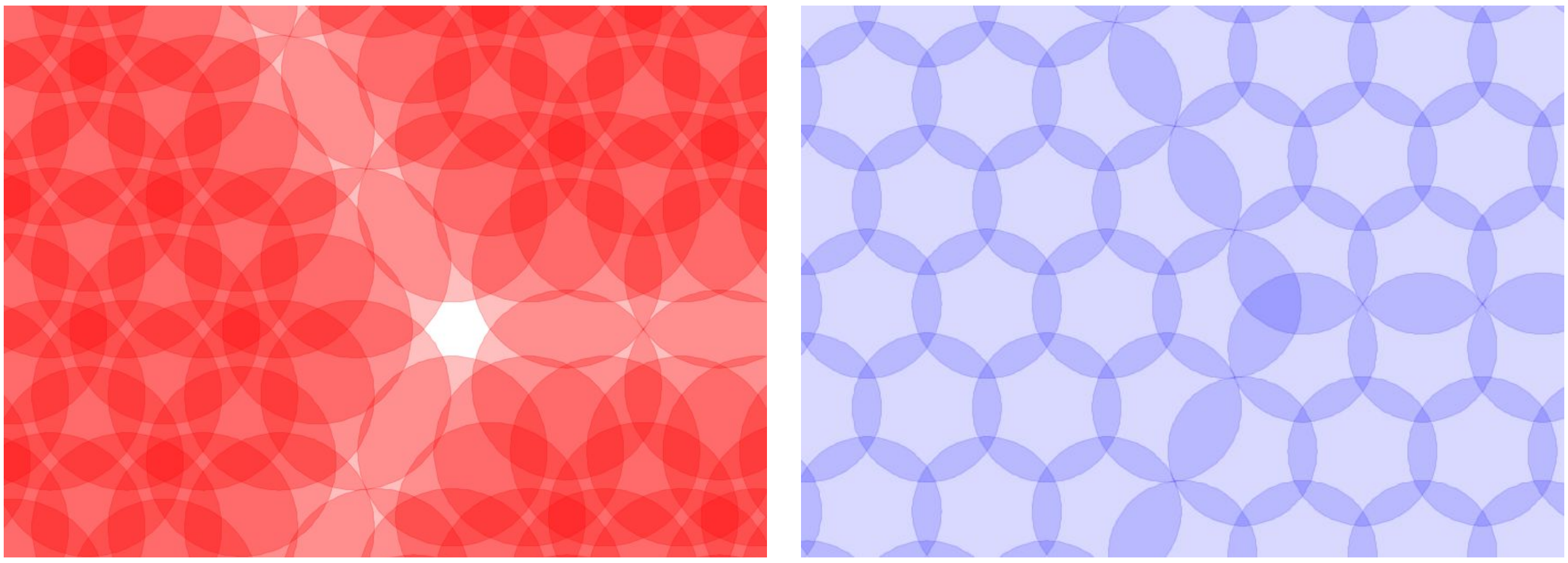

Figure 12: Field coverage (zoomed) for LMR fibers (red, at left) and HR fibers (blue, at right)

# 4. METROLOGY

## 4.1 Overview

The metrology system is vital to the operation of the Sphinx Fiber Positioner, providing feedback to control accurate re-positioning of the fiber feeds. The aim to measure all fiber positions in one exposure imposes challenging requirements on the metrology system. The CCD must have many pixels and the metrology goal corresponds to measuring image centroids to an accuracy equivalent to ~1/44$^{th}$ of a 6 µm pixel. With the camera positioned ~19m from the positioner, the corresponding angular accuracy goal is ~22 milliarcseconds, making the measurement very vulnerable to local seeing. This is a very challenging application, and to our knowledge, no multi-fiber positioners on telescopes are yet operating with full field metrology, although PFS (Subaru), 4MOST (VISTA), and DESI (Mayall) all have this intention.

The metrology system comprises of 2 major components:

- Metrology camera (including CCD and custom lens)
- Camera rotation unit

The metrology camera is mounted on a unit that is rotated synchronously with the instrument rotator holding the fiber positioner. With a fixed (non-co-rotating) metrology camera, the angular velocity at which the positioner is driven between science exposures would cause 1 second images of backlit spines to be too badly blurred, prohibiting accurate metrology for a large fraction of the time before the rotation ended. Limiting the exposure time would require very bright back illumination and exacerbate errors due to dome seeing. Utilising synchronous rotation avoids these issues and offers the additional advantage that the shorter side of the CCD can be √3/2 as large as if it had to accommodate the diagonal of the hexagonal field.

Key requirements for the metrology subsystem include:

- The metrology camera must measure the positions of some fiducials outside the spine array, including some on guide cameras.
- The error in measurement of fiber positions in normal mode is to be ≤ 3 µm with a goal of 2 µm.
- The goal for the error in measuring the final positions of fiber apertures is 1.5 µm.
- The hexagonal array of science spines will be ~582 mm corner to corner when spines are at their home positions.

## 4.2 Dome seeing

### 4.2.1 Spatial distribution

The best location for the metrology camera will be near the vertex of the primary mirror (see Figure 1 and Figure 3), where its central segment (1.44 m across corners) is missing. This results in the distance from the metrology camera lens to the front lens of the wide field corrector (WFC) being ~17 m. In this location, there will be very significant displacements of images arising from thermal inhomogeneity in the air traversed between the telescope top end and the metrology camera.

The best experimental evidence on the likely magnitude of the error due to seeing comes from measures made at Subaru for the PFS project by Siang-Yu Wang et al. Based on these measurements and our careful calculations, we conclude that the amplitudes of seeing-induced displacements will be small at spatial frequencies higher than ~10 cycle/m so that calibration with fiducials spaced apart by 100 mm or less will allow effective compensation.

### 4.2.2 In-field fiducials

By back-illuminating fiducials with a different wavelength to the spines, the fiducials can be located much closer to the spine pivots, ~200 mm away from the nominal focal surface, yet still maintain good focus to the metrology camera. This has the valuable advantage of not needing to sacrifice any science spines to accommodate these fiducials. This is made possible by the thinness of the Sphinx spines. To comfortably compensate for seeing-induced image motions, Sphinx will provide ~300 fiducials at a pitch of ~38mm.

Each spine module has 10 LEDs mounted on a PCB installed on the opposite side of the module to the piezo actuator PCB. The LEDs are small-footprint surface mount devices that are lit on every second module.

## 4.3 Metrology accuracy

Metrology software measures the precise location of all fibers at the focal surface, using the fiducials to account for scale/translation/distortion effects. It allows closed-loop control of the Fiber positioner. The speed and accuracy of the metrology software is a major factor in the overall speed and accuracy of the observation configuration.

The metrology software supports 3 modes of operation, with differing levels of accuracy and speed.

Table 5: Metrology accuracy

| Mode | Accuracy Req. (µm) | Accuracy Goal (µm) | Speed Req. (sec) | Speed Goal (sec) |
|---|---|---|---|---|
| Fast | 10.0 | 5.0 | 3 | 1 |
| Normal | 3.0 | 2.0 | 5 | 2 |
| Precise | 2.0 | 1.5 | 10 | 6 |

## 4.4 Metrology summary

The metrology camera mounted near the primary mirror vertex, using a 8956 x 6708 pixel CCD, combined with a custom lens design and special arrangements for calibration, is capable of achieving the very high accuracy needed for determining the positions of all fibers. To be sure that local seeing will not significantly degrade the measurement accuracy (with the metrology camera positioned ~17m from the vertex of the WFC) it is necessary to have many in-field fiducials with accurately predictable positions to calibrate the differential distortion due to local seeing. The fiber positioner provides ~300 in-field fiducials which are significantly offset from the focal surface (requiring a different wavelength of back-illumination to the science spines) to ensure that no science feeds need to be sacrificed. Co-rotating the metrology camera in sync with the fiber positioner ensures that the time overhead for fiber reconfiguration is minimised.

# 5. OBSERVING EFFICIENCY

## 5.1 Allocation efficiency

Saunders et al [6] undertook a study of the allocation yields resulting from various positioner geometries and target samples, for the DESI and 4MOST surveys. For DESI, dominated by weakly clustered sources with a surface density ~5 times the fiber density and observed 5 times over, the allocation efficiency was increased with tilting spines, overall by 5.7%, but much more (16%) for sparse targets requiring re-observation. For 4MOST, 85% allocation efficiency was achieved for simulated high-latitude samples in 6 passes, including 47% of targets in rich clusters.

In general, the allocation efficiency gain is largest when the target and fiber densities are about equal, and this condition is true for MSE at high latitudes. In this case, an allocation simulation for unclustered targets (including plausible exclusion radii) gives an efficiency of 84% for Sphinx vs 59% for a nominal phi-theta positioner, an efficiency gain of 43%. At higher target densities or at high resolution, the gain is less, but still significant. Any clustering of targets, on any scale, increases the advantages of Sphinx.

## 5.2 Small-scale clustering

The Sphinx actuator allows pairs of targets as close as 0.75 mm (7.0 arcseconds) to be observed, or triplets within any 1.06 mm (9.9 arcsecond) diameter circle. With multiple passes, it becomes possible to derive group and cluster velocity dispersions, a key MSE science goal.

## 5.3 Simultaneous LMR/HR observations

LMR and HR fibers are carried on separate spines, allowing complete sky coverage at HR and 3-fold sky coverage at LMR. No optical switches are needed to switch between HR and LMR use. This means that Sphinx can use all fibers, both LMR and HR, for all fields. At high latitudes, this means that ~1000 HR targets can always be observed, without interfering with LMR observations. For all purely stellar programs committed to both HR and LMR observations, it means up to 1000 extra LMR fibers are available for all fields.

## 5.4 Defocus

For each field observed, there is a distribution of spine tilts, and hence a range of axial positions of the tips. Since the telescope focus can only be set for one value, there is a defocus for most spines, causing an aperture efficiency loss for the starlight entering the fiber. The average defocus depends inversely on spine length, so the longest spines possible are preferred. The current prototype spine length is 250mm, but we are confident that a 300mm spine is viable, and are proposing to investigate this option further during the preliminary design phase.

The allocation software will minimize tilt where possible, so the average tilts will be smallest for LMR spines and for crowded fields, and largest for HR spines in uncrowded fields. For any given tilt distribution, there is an optimum focus, and the telescope focus would always be set to this optimum.

Saunders et al [6] determined the distribution of position radii for various 4MOST spine types and samples. There is some variation, but the RMS radius is always in the range 0.65-0.71 pitches.

Combining the tilt distributions with the injection efficiency then allows determination of the average and worst loss of injection efficiency due to defocus, as a function of spine length. For 250mm spines, the *worst* defocus losses are 13.7% and 8.7% for 0.8″ and 1.0″ fibers respectively. However, the *average* losses are just 1.3% for 0.8″ fibers and 0.9% for 1.0″ fibers. For 300mm spines, all these losses are reduced by 30%.

These are injection efficiency losses for point sources. For both faint galaxies and bright stars (MSE's 2 dominant source types), the efficiency loss is approximately equal to this injection efficiency loss. The efficiency loss is calculable for each target a priori, and so can be incorporated into the allocation strategy, to ensure adequate S/N is maintained.

### 5.5 Tilt losses

Spine tilt introduces a variable amount of geometric FRD into the light entering the fibers. This propagates through the fibers into the beam exiting the fiber, increasing the collimator overfilling losses. For MSE, the effect is partially mitigated because the telescope beam is already apodized by the hexagonal primary.

We have modelled the simulated beams exiting the fibers, with various amounts of input tilt, and with FRD as measured by Zhang et al [7] for the high NA fibers proposed for MSE (Figure 13). The additional collimator losses due to spine tilt depend on collimator speed, wich differs for LMR and HR. The maximum tilt losses are 7.9% at LMR and 5.3% for HR, while the average losses are 3.1% at LMR and 1.5% at HR. This is for a fiber at the center of the MSE field, the losses are smaller elsewhere because the effective telescope speed is slower, principally due to vignetting [8]. The additional losses for 300mm spines would be ~30% less.

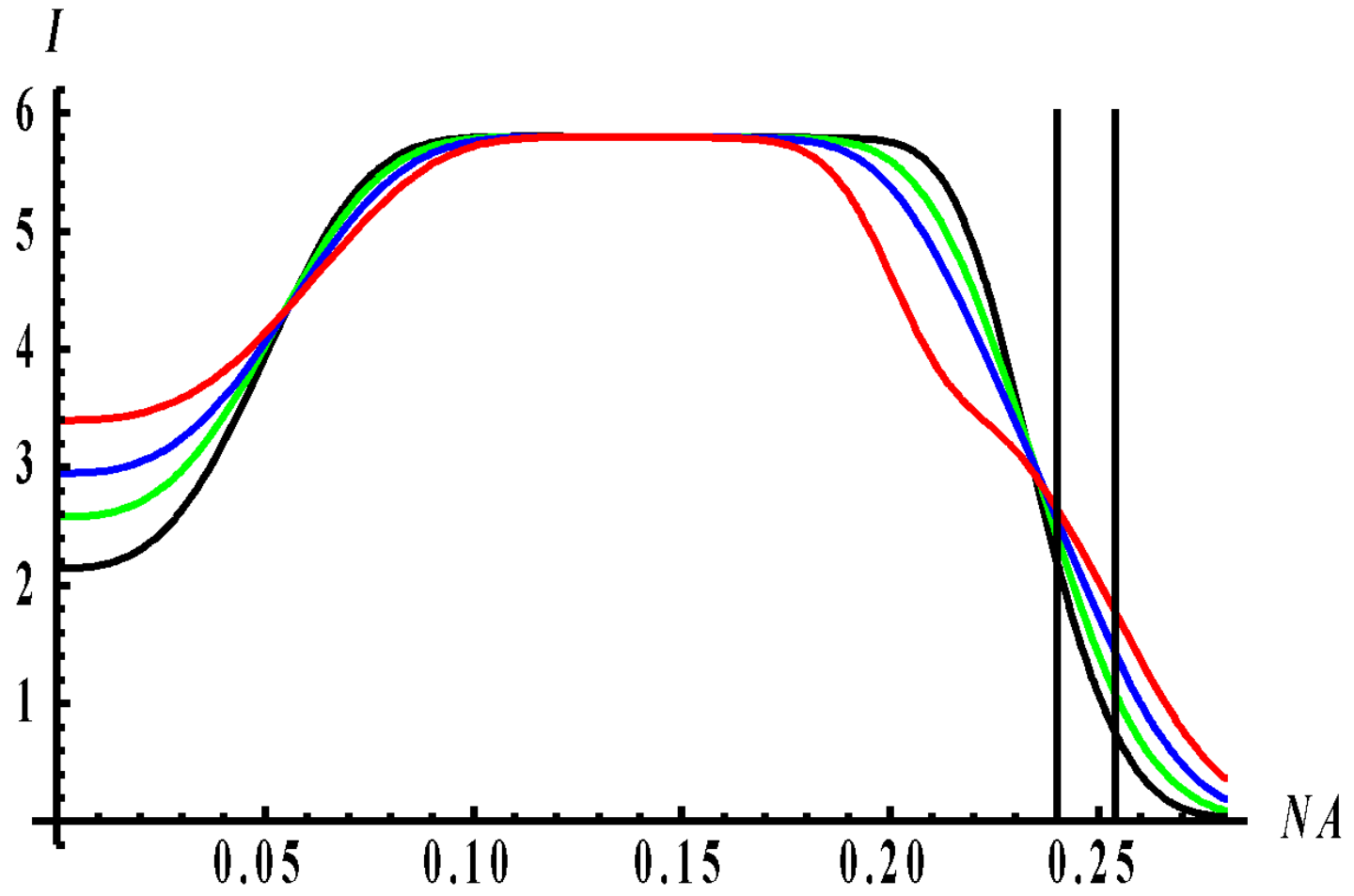

Figure 13: Radial profiles of the beams exiting the fibers, for spine tilts of 0 (black), 1° (green, 1.5° (blue), and 2° (red). Vertical lines show the nominal collimator acceptance speeds for LMR (f/2.08, left) and HR (f/1.97, right).

In crowded fields, the tilts and resulting losses are somewhat smaller, with the losses varying roughly as the square of the tilt.

In all cases, the tilt losses are calculable for each target *a priori,* and so can be incorporated into the allocation strategy, to ensure adequate S/N is maintained.

In principle, spine tilt could lead directly to increased attenuation losses within the fiber. However, Zhang et al [7] found no evidence for such losses, even at tilts much larger than proposed here.

### 5.6 PSF variations

For any spectrograph, the PSF at any wavelength is variable along the slit. This is undesirable, but can be calibrated and corrected for by the arcs and fiber-flats. Spine tilt introduces a further effect, which is that the variations in the far-field beam exiting the fibers change the apodization pattern of the beam in the spectrographs. The speed-dependent nature of the spectrograph aberrations then causes tilt-dependent variations in the PSF on the detector, for the same fiber and wavelength, from one field to the next.  In principal, this could impact on the spectral extraction from 2D to 1D, and on the sky subtraction and radial velocity precision.

These variations have been calculated directly for the LMR spectrograph design. The maximum changes in the PSF due to tilt are ~0.3% of the peak height, and the maximum centroid shift is ~0.2% of the projected fiber diameter (or 1/125 pixel). They are several times smaller than was found for DESI and 4MOST, because the maximum tilt is a smaller fraction of the telescope NA, and the PSF changes vary as the square of this fraction.

These variations are an order of magnitude smaller than those in any case arising due to varying spectrograph aberrations along the slit, which are ~5% of peak height. So spine tilt increases the total PSF variation (the two effects added in quadrature) by <0.2%.

So spine tilt has virtually no effect on data quality, as long as arc and fiber-flat calibration frames have the same PSF as the data, for every fiber and for every field. Since arcs and flats will be taken immediately before and/or after the data frames, they will always have the same spine tilts and hence the same PSFs, limited only by the ability of the calibration system to mimic the telescope pupil, which is the same for any positioner system. Hence spine tilt causes no additional error in the extractions.

In principle, spine tilt could cause *near-field* changes in the light emitted by the fibers, which translate directly to the PSF. However, Zhang et al [7] found no evidence for such changes at the 0.5% level in the fibers proposed for MSE; and in any case, the same changes would apply to the calibration frames in the same way as for the far-field, and again there is no error arising in the extractions.

### 5.7 Observing efficiency summary

Sphinx offers large allocation efficiency gains – at least 40% - for important classes of LMR observations: LMR fields observed only once, or targets that must be re-observed on multiple visits (whether for time-series or S/N reasons), or clustered targets. For multiple sky passes with unclustered targets each observed just once, the gains are ~5%.

For HR, the allocation efficiency gain is much more marginal, 0-7% depending on the number of passes. However, the optical switches are avoided, giving a significant efficiency gain.

For both LMR and HR, there is a significant efficiency gain – of up to 33% - arising from simultaneous LMR/HR observations with all fibers.

Spine defocus incurs a small *average* efficiency penalty, ~1-3% depending on fiber size, target brightness and spine length. Individual sources will suffer larger, but predictable, losses (up to 6-27% depending on target brightness, fiber size and spine length).

Average additional collimator overfilling losses are 1-3%, depending on fiber size and spine length; worst affected spines will suffer 4-9% losses.

There are negligible efficiency losses from attenuation differences or PSF variations.

Overall, Sphinx offers greatly improved efficiency for many classes of LMR observations; marginal average gains for all other classes of LMR target, and marginal average gains for HR targets. These gains are at the cost of a significant, but predictable, efficiency variation between targets. If needed, the allocation software could include a penalty against allocating faint targets at large tilts, to ensure required minimum S/N is always maintained.

## 6. CONCLUSION

The concept design for the Sphinx Fiber Positioner is a very low technical risk solution for MSE, due to its technical simplicity, design maturity, proven performance, low maintenance requirements, and AAO's long history of instrumentation experience. The tilting spine design offers significant efficiency advantages including: large and overlapping patrol areas, small exclusion radii, simultaneous full-capability LMR and HR use, and no optical switches. For LMR observations, the increased allocation efficiency always outweighs the efficiency losses per fiber, by large amounts for clustered sources or those requiring observation on every pass of the sky. For HR observations, the efficiency gain from avoiding optical switches outweighs any tilt losses.

The slimness of the spine design in conjunction with a unique metrology system allows large numbers of fiducials to correct the metrology for dome seeing, without sacrificing any science fibers.

As we enter the preliminary design phase of this project, we are confident that Sphinx will be a very successful part of the MSE instrument.